\documentclass[preprint, 1p]{elsarticle}

\usepackage{amssymb}
\usepackage{amsmath}
\usepackage{multicol}
\usepackage[utf8]{inputenc}
\usepackage[T1]{fontenc}
\usepackage[colorlinks=true, allcolors=blue]{hyperref}
\hypersetup{
        pdfauthor          = {Julius Römer},
        pdftitle           = {Radiation Hardness of a Wide Spectral Range SiPM with Quasi-Spherical Junction},
        pdfsubject         = {NDIP20 Proceedings},
        pdfkeywords        = {TAPD, SiPM, radiation damage},
        urlcolor           = red,
        citecolor          = blue,
        linkcolor          = blue,
        pdfborder          = {1 1 1}
        }
\usepackage{lineno}

\usepackage{float}
\usepackage{subfigure}
\usepackage{caption}
\usepackage{multirow}
\usepackage{bm}
\usepackage{siunitx}
\journal{NIM Proceedings}

\begin{document}

\begin{frontmatter}



\title{Radiation Hardness of a Wide Spectral Range SiPM\\with Quasi-Spherical Junction}


\author[uhh]{Julius~Römer\corref{cor1}}
\author[uhh]{Erika~Garutti}
\author[inst1,inst2]{Wolfgang~Schmailzl}
\author[uhh]{Jörn~Schwandt}
\author[uhh]{Stephan~Martens}

\cortext[cor1]{Corresponding author. E-mail: julius.roemer@desy.de}

\affiliation[uhh]{organization={Institut~für~Experimentalphysik, Universität~Hamburg},
            addressline={Luruper~Chaussee~149}, 
            city={Hamburg},
            postcode={22607}, 
            state={Hamburg},
            country={Germany}}

\affiliation[inst1]{organization={Broadcom~Inc.},
            addressline={Wernerwerkstrasse~2}, 
            city={Regensburg},
            postcode={93049}, 
            state={Bayern},
            country={Germany}}

\affiliation[inst2]{organization={\mbox{Universität~der~Bundeswehr~München}},
            \protect\\addressline={\mbox{Werner-Heisenberg-Weg~39,}},
            city={Neubiberg},
            postcode={85577}, 
            state={Bayern},
            country={Germany}}


\begin{abstract}
New pixel geometries are on the rise to achieve high sensitivity in near-infrared wavelengths with silicon photomultipliers (SiPMs). We test prototypes of the tip avalanche photo-diodes, which feature a quasi-spherical \mbox{p-n} junction and a high photodetection efficiency over a wide spectral range, and analyze the performance after neutron irradiation. The observed increase in dark count rate is significantly smaller than for a SiPM with a conventional design, indicating a good radiation hardness of the pixel geometry.
\end{abstract}



\end{frontmatter}


\section{Introduction}
\label{sec:intro}
Silicon photo-multipliers (SiPMs) have a wide range of applications, such as medical imaging \cite{bisogniMedicalApplicationsSilicon2019}, particle physics experiments \cite{simonSiliconPhotomultipliersParticle2019}, imaging in space \cite{casolinoSiPMDevelopmentSpaceborne2021}, or light detection and ranging (LIDAR) systems \cite{agishevLidarSiPMCapabilities2013}. Many efforts in the development of SiPMs target the optimization of photo-detection efficiency in combination with additional requirements for specific applications, such as radiation hardness for space and high-energy experiments or high red-sensitivity and dynamic range for LIDAR systems \cite{acerbiSiliconPhotomultipliersTechnology2019}. The realization of SiPMs with enhanced sensitivity to near-infrared light is challenging due to the large penetration depth of photons in silicon at these wavelengths. For these applications, a larger photoelectron collection depth is required. With a conventional design built around a planar \mbox{p-n junction}, a limit is reached as the effective active area decreases with increased depletion depth \cite{acerbiSiliconPhotomultipliersSinglephoton2018}.\\
To overcome this limitation without sacrificing dynamic range, several attempts with a novel pixel geometry have been made \cite{vansieleghemBacksideIlluminatedChargeFocusingSilicon2022, engelmannTipAvalanchePhotodiode2021}. We examine prototypes of the tip avalanche photo-diode (TAPD) SiPM presented in \cite{engelmannTipAvalanchePhotodiode2021}. This device features a quasi-spherical \mbox{p-n} junction embedded at a few micrometers below the silicon surface. This geometry provides a homogeneous electric field distribution around the tip and a focusing effect regarding the charge collection.  For a pixel pitch of \SI{15}{\micro\metre}, a recovery time constant of $\tau=\SI{4.5}{ns}$ and record photodetection efficiency ($PDE$) values of \SI{73}{\%} at \SI{600}{nm} and \SI{22}{\%} at \SI{900}{nm} have been reported~\cite{engelmannTipAvalanchePhotodiode2021}.\\
\begin{figure}[h]
    \centering
    \includegraphics[width=0.5\textwidth]{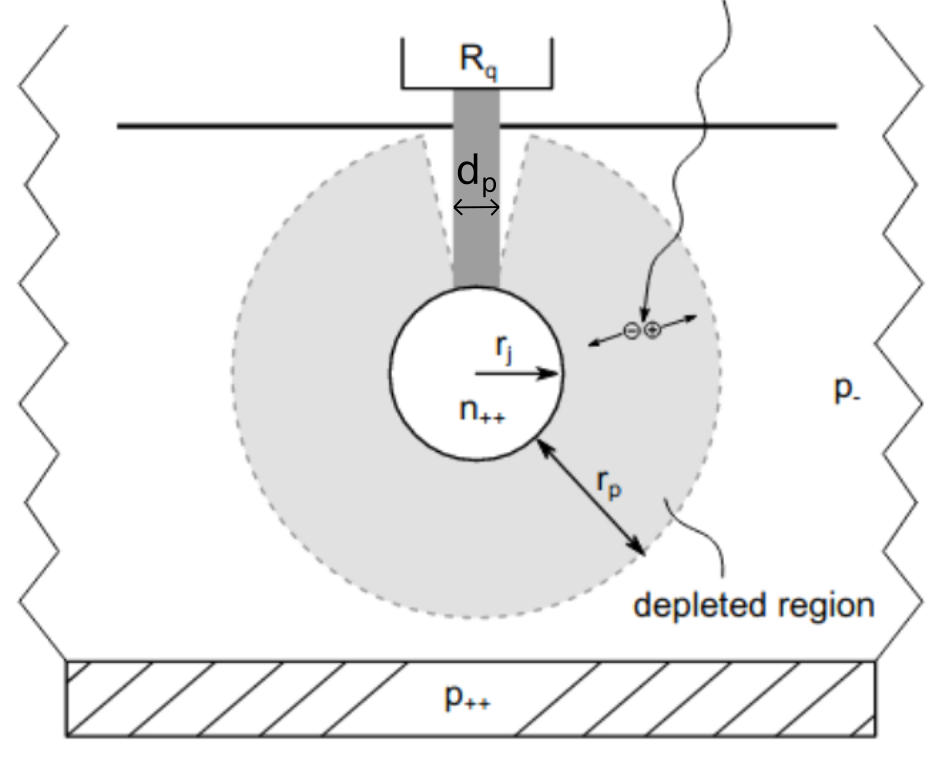}
    \caption{Sketch of the cross-section of the TAPD. The bias is provided via the quenching resistor on the surface through the pillar to the n-region. The pillar diameter $d_{\text{p}}$ is the nominal value given to characterize the size $r_j$ of the n-region. Taken from~\cite{engelmannTipAvalanchePhotodiode2021}.}
    \label{fig:sketch}
\end{figure}

We characterize the device and investigate its performance after neutron irradiation. We were provided with samples varying in the size of the n-doped region, represented by the nominal pillar diameter $d_{\text{p}}$ as shown in Fig.~\ref{fig:sketch}. The SiPM features an array of \SI{5016}{pixels} in an area of \SI{1}{mm^2}. Devices with a pillar diameter of \SIlist{0.6;0.8;1.0}{\micro\metre} were studied. Irradiation with neutrons with \SI{1}{MeV} neutron equivalent fluences $\Phi_{\text{eq}}$ of \SIlist{1e10;1e11;1e12}{cm^{-2}} has been performed at the TRIGA nuclear reactor at the Jožef Stefan Institute in Ljubljana\footnote{\url{https://ric.ijs.si/en/splosna-predstavitev/}}.
Neutron irradiation can introduce a reduction in signal-to-noise ratio due to trapping of charge carriers, a shift in breakdown voltage due to a change in effective doping concentration, changes in quenching resistance and most significantly, a dramatic increase in dark count rates ($DCR$) \cite{garuttiRadiationDamageSiPMs2019}. The activity of SiPM pixels due to dark counts can result in a loss of $PDE$ \cite{garuttiCharacterisationHighlyRadiationdamaged2017}.\\
In this work, we compare our results for the TAPD-SiPM with a conventional KETEK MP15 SiPM with 4384 pixels á \SI{15}{\micro\metre} pixel pitch, which was investigated in \cite{vignaliNeutronInducedRadiation2016}.

\section{Analysis of charge spectra}
To evaluate the charge spectra, the SiPM is wire-bonded on a ceramic with pin connectors, which is connected to a trans-impedance amplifier with an amplification of $G_{\text{TIA}}=\SI{3000}{\Omega}$. The device is illuminated with a low-intensity laser pulse with a wavelength of \SI{675}{nm} and a pulse length of \SI{35}{ps} full-width half-maximum. The laser diode is driven at a frequency of \SI{100}{kHz} and the intensity of the light is tuned with neutral density filters. The outgoing signal is recorded with a Rohde + Schwarz RTO 2044 oscilloscope and the integral of the area beneath the signal of \SI{250000}{events} is extracted to create charge histograms. The gain $G$ is then evaluated as the difference between adjacent photo-electron peak positions in the charge spectrum. The turn-off voltage $V_{\text{off}}$ is determined by extrapolating the linear gain-voltage dependence to $G=0$ \cite{chmillStudyBreakdownVoltage2017}.

\begin{figure}[H]
\begin{subfigure}
    \centering
    \includegraphics[width=0.49\linewidth]{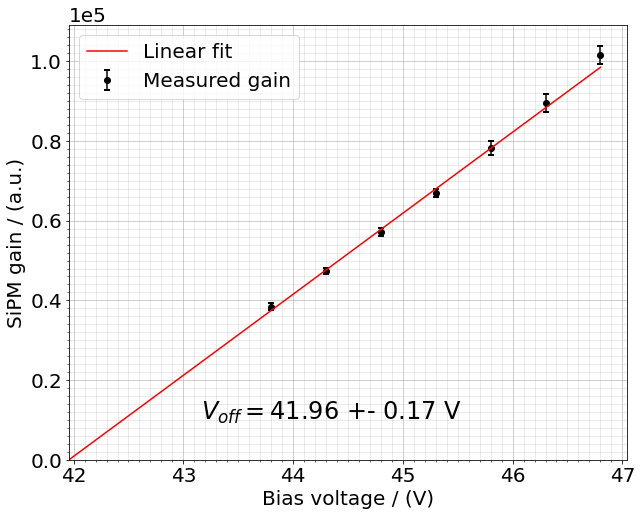}
\end{subfigure}
\begin{subfigure}
    \centering
    \includegraphics[width=0.49\linewidth]{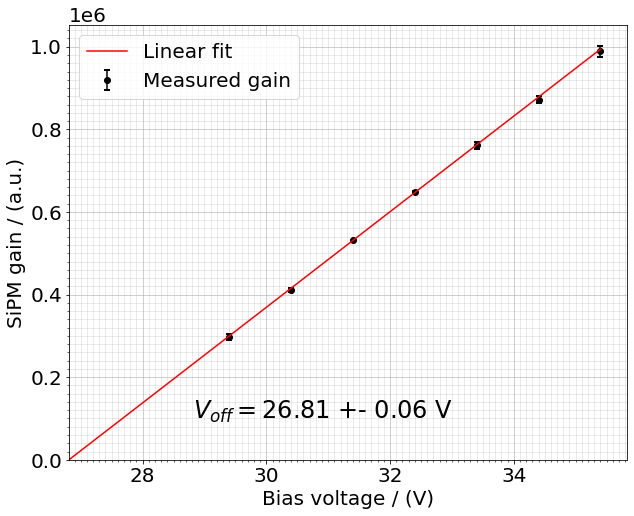}
    \end{subfigure}
    \caption{SiPM gain versus bias voltage for the TAPD \SI{0.6}{\micro\metre} (left) and the MP15 (right). A linear fit is applied to determine $V_{\text{off}}$. It is noted that the TAPD measurements are conducted at \SI{24.9}{\degree C} and for the MP15 at \SI{21.9}{\degree C}. To compare the values of $V_{\text{off}}$, we can use $\mathrm{d}V_{\text{BD}}/\mathrm{d}T$.}
        \label{fig:offtapd}
\end{figure}

The gain-voltage behavior of both devices is shown in Fig.~\ref{fig:offtapd}. The TAPD gain is roughly a factor of 10 lower than that of the MP15. As the charge measurement setup is not temperature-controlled, the measurement has been conducted at \SI{24.9}{\degree C} for the TAPD \SI{0.6}{\micro\metre} and at \SI{21.9}{\degree C} for the MP15. To compare the breakdown voltage, which is shown in section~\ref{sec:breakdown}, and the turn-off voltage, we add the product of the temperature difference and $\mathrm{d}V_{\text{BD}}/\mathrm{d}T$ as shown in table~\ref{tab:v_bd}. After correcting for the temperature difference between the current and the charge measurements, we found that $V_{\text{off}}$ is \SI{0.72\pm0.06}{V} lower than $V_{\text{BD}}$ for the non-irradiated MP15 sample, but up to \SI{0.38\pm0.14}{V} larger for the non-irradiated TAPD sample. The relationship $V_{\text{off}}>V_{\text{BD}}$ is unexpected, and a slightly non-linear $G(V)$-behavior observed for the TAPD might be the reason. Further investigations are planned.

\section{Analysis of current-voltage measurements}
\label{sec:meas}
In the $I$--$V$ measurement setup, the sample is placed on a temperature-controlled cold chuck. The prototype is contacted with needles and shielded electromagnetically with a grounded, light-tight box. A Keithley 6517B is used to supply the bias voltage and to measure the current. Measurements have been conducted in forward and reverse bias at \SI{-30}{\degree C} and \SI{20}{\degree C}. In reverse bias, measurements are performed in the dark ($I_{\text{dark}}$) and under the illumination of a diffused LED ($I_{\text{light}}$) at a wavelength of \SI{470}{nm}.

\subsection{Quenching resistance}
The measurements are conducted in forward bias to extract the quenching resistance $R_{\text{q}}$ \cite{klannerCharacterisationSiPMs2019b} using 
\begin{equation}
    R_{\text{q}} = N_{\text{pix}}\cdot \left(\frac{\mathrm{d}I_{\text{f}}}{\mathrm{d}V}\right)^{-1} \label{eq:rq}\,,
\end{equation}
where $N_{\text{pix}}$ is the number of pixels, $I_{\text{f}}$ the measured current at applied forward biasing voltage $V$.\\

While the measurements of the 1 mm$^2$-array are used to detect a shift of $R_{\text{q}}$ with irradiation, only the measurements of single pixel structures showed the linear behavior required to apply Eq.~\ref{eq:rq} for retrieving the absolute value. Applying this method, a quenching resistance of \SI{615 \pm 35}{\kilo\ohm} has been measured. While no significant increase of $R_{\text{q}}$ has been observed for most fluences, the highest fluence caused an increase of approx. \SIrange{1.6}{5.8}{\%}. This is consistent with previous results for the MP15. Between \SI{-30}{\degree C} and \SI{20}{\degree C}, a linear temperature dependence value of \SI{-3.37 \pm 0.05}{\kilo\ohm \per K} has been determined for the TAPD.

\subsection{Breakdown voltage}\label{sec:breakdown}
Similar to \cite{klannerCharacterisationSiPMs2019b, chmillStudyBreakdownVoltage2017}, we determine the breakdown voltage as the voltage where the inverse logarithmic derivative $ILD$ has its minimum, where
\begin{equation}
    ILD = \left(\frac{\mathrm{d}\ln I}{\mathrm{d}V}\right)^{-1}\,.\label{eq:ild}
\end{equation}
The minimum was evaluated using a parabolic fit and a cubic spline fit applied to the raw data of $ILD$.\\
\begin{figure}[H]
\begin{minipage}[t]{0.48\textwidth}
    \centering
    \vspace{-2.75cm}
    \begin{tabular}{|c|c|c|}
    \hline
    \multirow{ 2}{*}{Device} &  $\bm{V_{\textbf{BD}}}$ \SI{20}{\degree C}  & $\bm{\mathrm{d}V_{\textbf{BD}}/\mathrm{d}T}$ \\
    &/ (\SI{}{V}) &/ (\SI{}{mV\per K})\\\hline \hline
    MP15 &\SI{27.46 \pm 0.03}{} & \SI{20.9 \pm 0.2}{}\\ \hline
    TAPD \SI{0.6}{\micro\metre} & \SI{41.81 \pm 0.08}{} & \SI{25.7 \pm 0.2}{}\\ \hline
    TAPD \SI{0.8}{\micro\metre} & \SI{46.18 \pm 0.09}{} & \SI{28.3 \pm 0.1}{}\\ \hline
    TAPD \SI{1.0}{\micro\metre} & \SI{48.98 \pm 0.06}{} & \SI{29.6 \pm 1.2}{}\\\hline 
    \end{tabular}
    \captionof{table}{The mean breakdown voltage as of all samples before irradiation and the mean temperature dependence measured in the range of \SIrange{-30}{20}{\degree C}.}
    \label{tab:v_bd}
\end{minipage}
\hspace{0.04\textwidth}
\begin{minipage}[t]{0.48\textwidth}
    \centering
    \includegraphics[width=\textwidth]{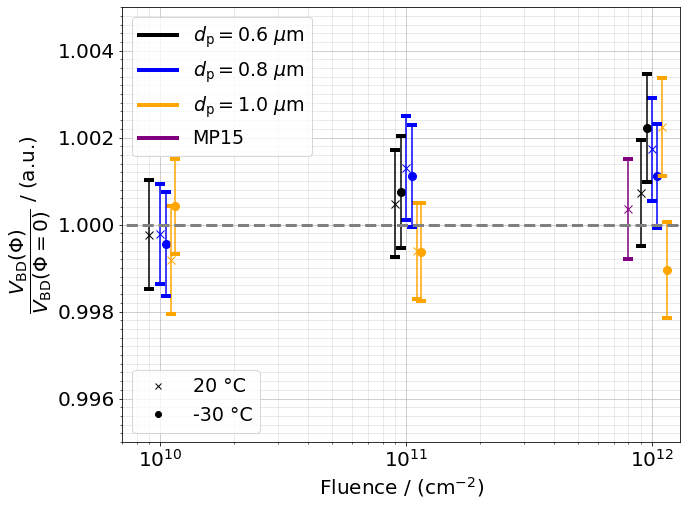}
    \caption{Observed change in breakdown voltage after irradiation. The markers have been separated horizontally for increased visibility. While the TAPD with a pillar diameter of \SI{1.0}{\micro\metre} at low temperatures shows a decline, the other TAPD samples show an increase of  \SI{0.16\pm0.06}{\%} at the highest fluence.}
    \label{fig:bd_shift}
\end{minipage}
\end{figure}

The breakdown voltage for the compared devices at \SI{20}{\degree C} and the temperature dependence of the breakdown voltage $\mathrm{d}V_{\text{BD}}/\mathrm{d}T$ are given in table~\ref{tab:v_bd}. Since the electric field strength at the sphere edge scales with $1/r^2$, $V_{\text{BD}}$ increases with growing pillar size. While the shift with temperature is larger for the TAPD than for the MP15, it is comparable to commercially available devices.
The ratios of the breakdown voltage before and after irradiation are displayed in Fig.~\ref{fig:bd_shift}. While the breakdown voltage of the TAPD with a pillar diameter of \SI{1.0}{\micro\metre} at low temperatures shows a decline with increasing fluence, the other TAPD samples show an increase of \SI{0.16\pm0.06}{\%} at the highest fluence. For the MP15, no change in breakdown voltage is observed.

\subsection{Photocurrent}
To investigate the response to light, we use the photocurrent $I_{\text{photo}}=I_{\text{light}}-I_{\text{dark}}$ and normalize it to 
\begin{equation}
    I_{\text{photo}}^{\text{norm}} = \frac{I_{\text{photo}}}{R_\gamma \cdot q_0}\,,\label{eq:photonorm}
\end{equation}
where $R_\gamma = I_{\text{photo,\,G=1}}/q_0$ is the rate of generated photoelectrons, extracted at $I_{\text{photo,\,G=1}}\approx I_{\text{photo}}(V_{\text{bias}}=10\,\mathrm{V})$.
We can model this normalized photocurrent with
\begin{equation}
    I_{\text{photo}}^{\text{norm}} = G\cdot PDE \cdot (1+CN)\,. \label{eq:photomodel}
\end{equation}
A comparison of $I_{\text{photo}}^{\text{norm}}$ before and after irradiation indicates whether there has been a change in the parameters in Eq.~\ref{eq:photomodel}.\\
\begin{figure}[h]
\begin{minipage}[t]{0.49\textwidth}
    \centering
    \includegraphics[width=\textwidth]{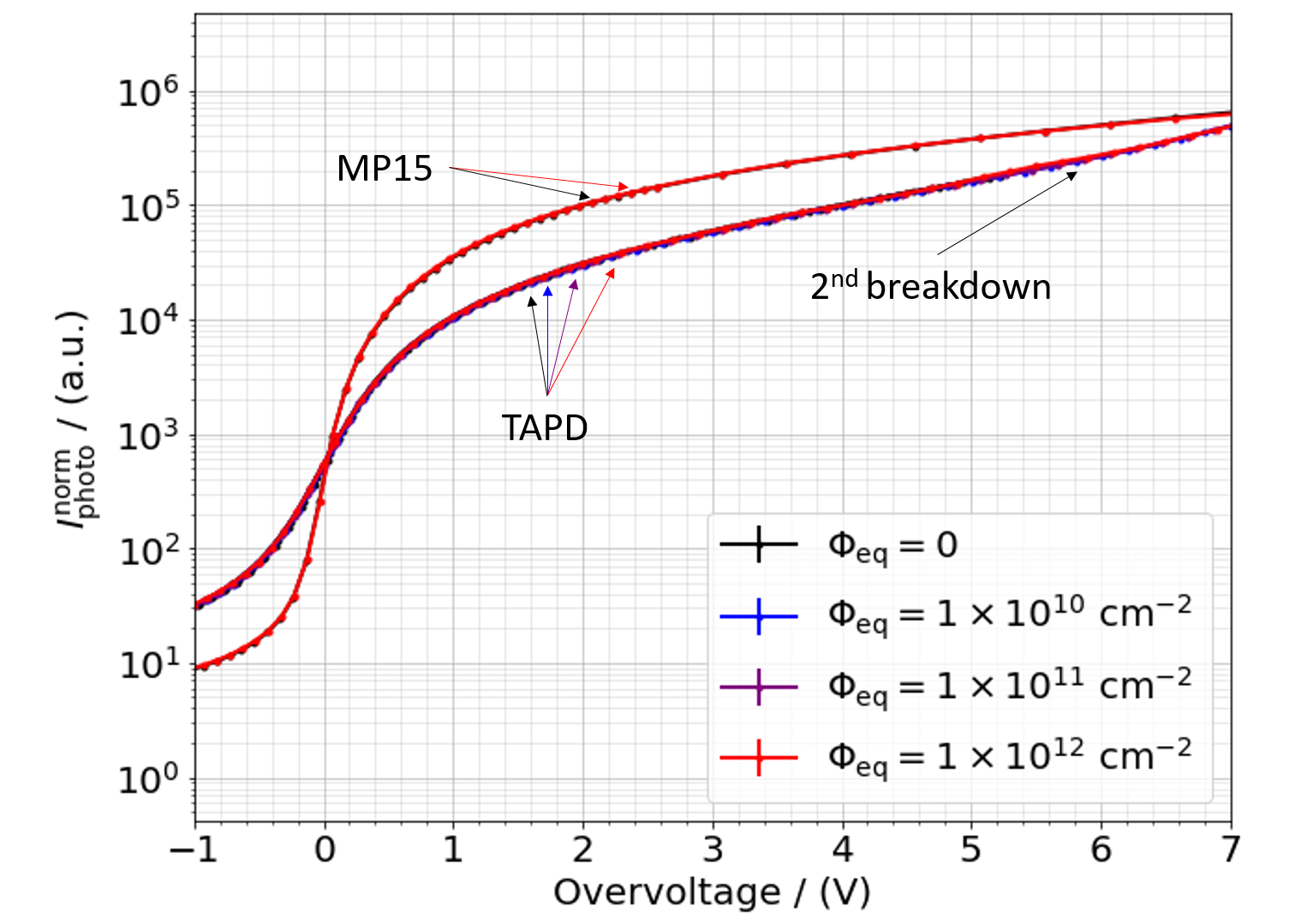}
    \caption{The normalized photocurrents of the TAPD \SI{0.6}{\micro\metre} and the MP15 at \SI{20}{\degree C}. For the TAPD, the operation range is limited to roughly \SI{5}{V} overvoltage, due to the second breakdown.}
    \label{fig:normphoto}
    \end{minipage}
    \hspace{0.02\textwidth}
\begin{minipage}[t]{0.49\textwidth}
    \centering
    \includegraphics[width=\textwidth]{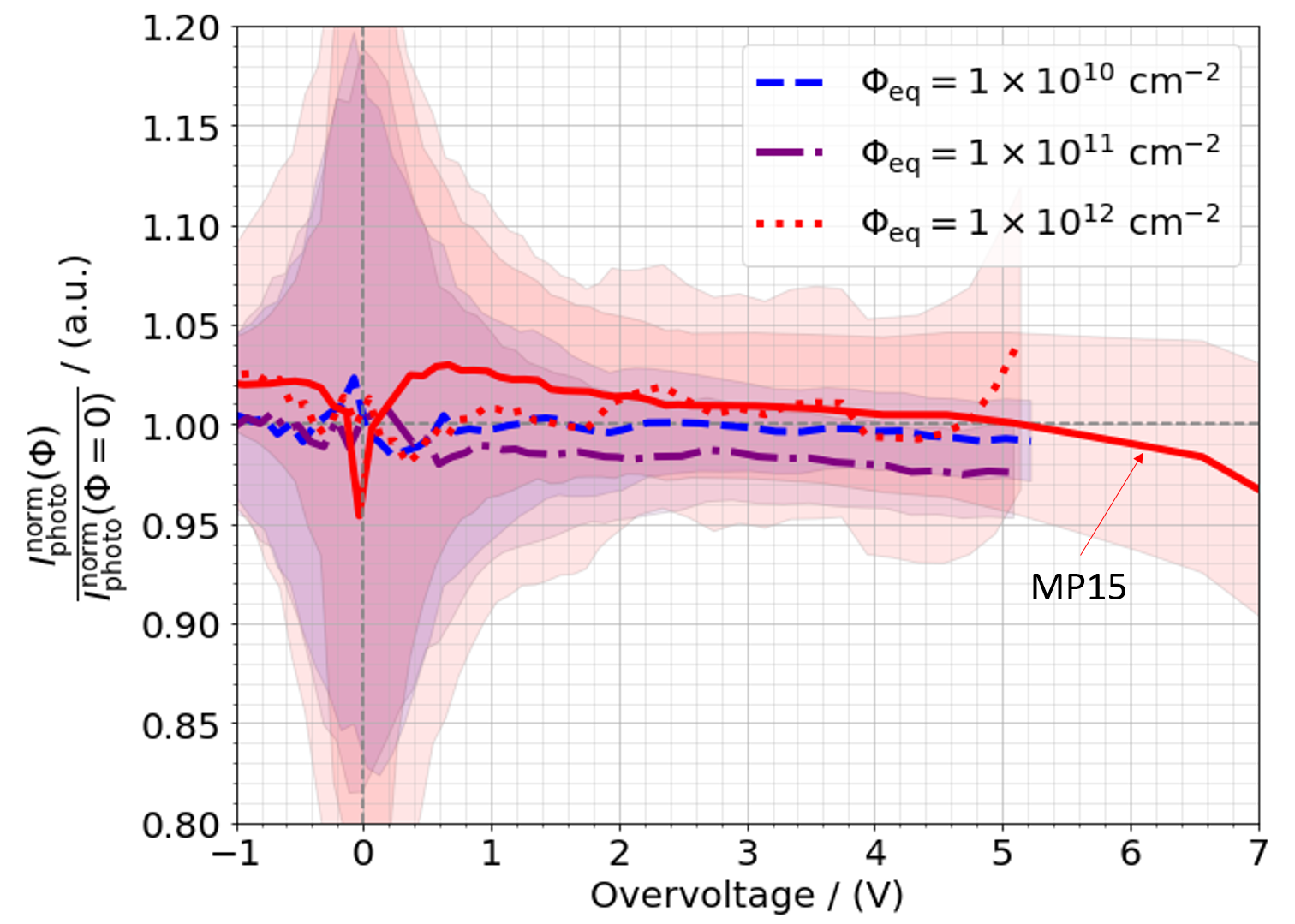}
    \caption{The ratio between the normalized photocurrents before and after irradiation for the TAPD \SI{0.6}{\micro\metre} and the MP15 at \SI{20}{\degree C}. The ratio is equal to 1 within standard errors. Therefore no change in photocurrent was observed. The reduction for the MP15 at high voltage could be due to the pixel occupancy due to dark counts, which reaches a few percent.}
    \label{fig:normphotoratio}
        \end{minipage}
\end{figure}
The normalized photocurrent in the operation range of the TAPD is shown in Fig.~\ref{fig:normphoto}. At an overvoltage $V_{\text{over}}=V_{\text{bias}}-V_{\text{BD}}>\SI{5}{V}$, the current increases more rapidly, indicating that the second breakdown and therefore the end of the operation range is reached. As seen in Fig.~\ref{fig:normphotoratio}, the ratio of the normalized photocurrents before and after irradiation $I_{\text{photo}}^{\text{norm}}(\Phi)/I_{\text{photo}}^{\text{norm}}(\Phi=0)$ is equal to one within standard errors. Therefore, no change in photocurrent within the operation range of the TAPD has been observed. 

\subsection{Dark current}
To estimate the $DCR$ and the pixel occupancy due to dark counts $\eta_{\text{DC}}$, we use a model for $I_{\text{dark}}$ described in \cite{garuttiCharacterisationHighlyRadiationdamaged2017}:
\begin{equation}
    \frac{I_{\text{dark}}}{G\cdot q_0} =DCR\cdot (1 + CN) =\frac{N_{\text{pix}}\cdot \eta_{\text{DC}}}{\Delta t}\,,\label{eq:idark}
\end{equation}
where $G$ is the gain, $q_0$ is the elementary charge, $CN$ the probability of correlated noise such as afterpulses and crosstalk, and $\Delta t$ is the pixel recovery time after a Geiger discharge.
We assume $G\cdot q_0 =(V_{\text{bias}}-V_{\text{off}})\cdot (C_{\text{pix}}+C_{\text{q}})$, where $V_{\text{off}}$ is the turn-off voltage  \cite{chmillStudyBreakdownVoltage2017}, $V_{\text{bias}}$ is the applied bias voltage, $C_{\text{q}}$ is the capacitance of the pixel and $C_{\text{q}}$ is the parasitic capacitance of the quenching resistor. For the recovery time, we assume $\Delta t = (C_{\text{pix}}+C_{\text{q}}) \cdot R_{\text{q}}$. This enables us to calculate $\eta_{\text{DC}}$ after rearranging Eq.~\ref{eq:idark}
\begin{equation}
    \eta_{\text{DC}} = \frac{I_{\text{dark}}\cdot \Delta t}{G\cdot q_0 \cdot N_{\text{pix}}} = \frac{I_{\text{dark}}\cdot R_{\text{q}}}{(V_{\text{bias}}-V_{\text{off}})\cdot N_{\text{pix}}}\,. \label{eq:eta}
\end{equation}
With knowledge of the recovery time, we can also calculate an estimate for the DCR using
\begin{equation}
    DCR \cdot (1+CN) = \frac{I_{\text{dark}}\cdot R_{\text{q}}}{(V_{\text{bias}}-V_{\text{off}})\cdot \Delta t}\,.\label{eq:dcr}
\end{equation}
The turn-off voltage has been determined using charge histograms, while the recovery time constant for the TAPD of $\tau=\SI{4.5}{ns}$ is taken from \cite{engelmannTipAvalanchePhotodiode2021} and  for the MP15 $\tau=\SI{11}{ns}$ as described in \cite{nitschkeCharacterizationSiliconPhotomultipliers2016}.\\

\begin{figure}[h]
\begin{minipage}[t]{0.49\textwidth}
    \centering
    \includegraphics[width=\textwidth]{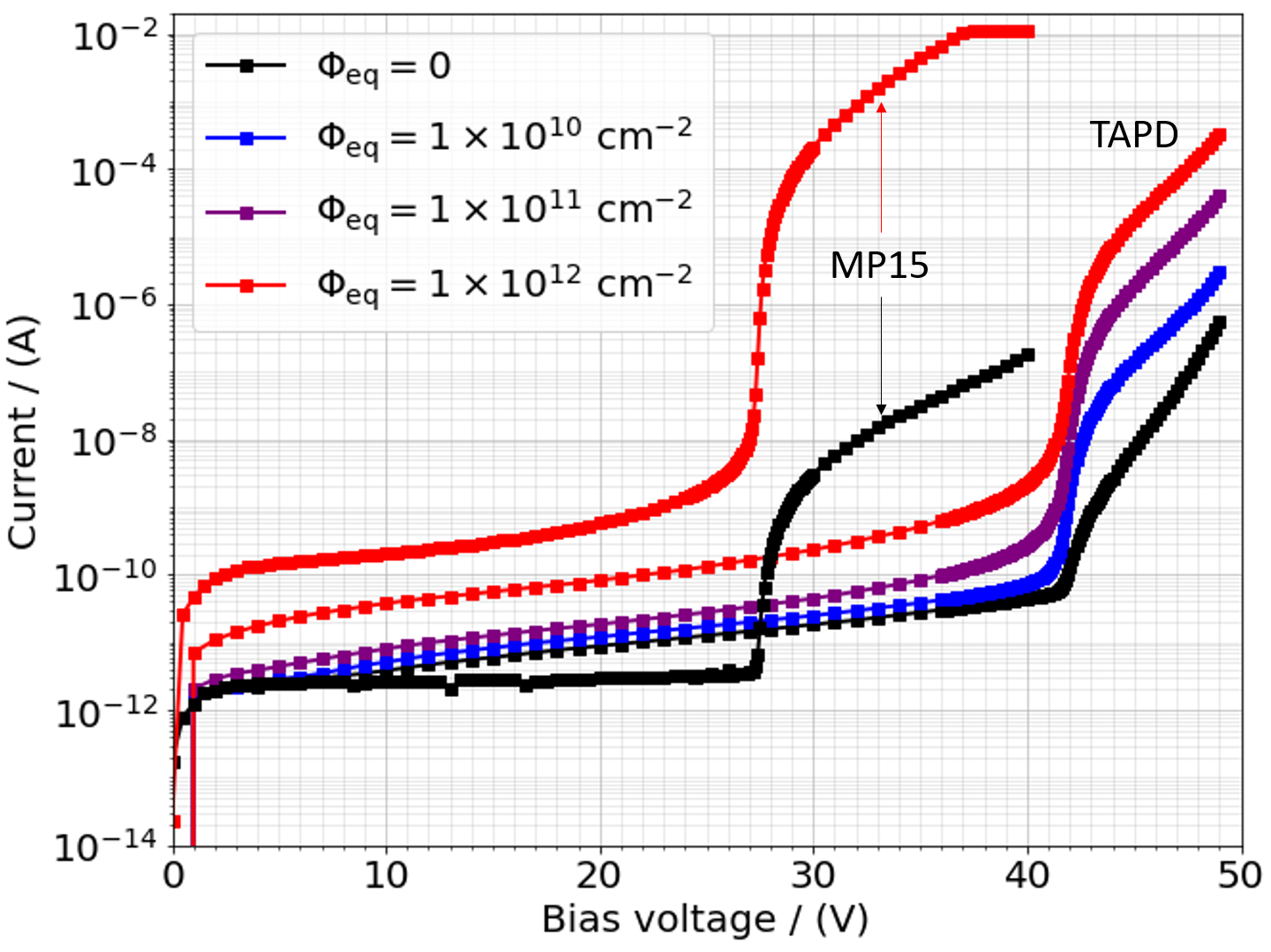}
    \caption{$I_{\text{dark}}$ at \SI{20}{\degree C} for the TAPD \SI{0.6}{\micro\metre} and the MP15. At an overvoltage of \SI{10}{V}, the MP15 current reaches the safety current limit of \SI{10}{mA}.}
    \label{fig:idark}
    \end{minipage}
    \hspace{0.02\textwidth}
\begin{minipage}[t]{0.49\textwidth}
    \centering
    \includegraphics[width=\textwidth]{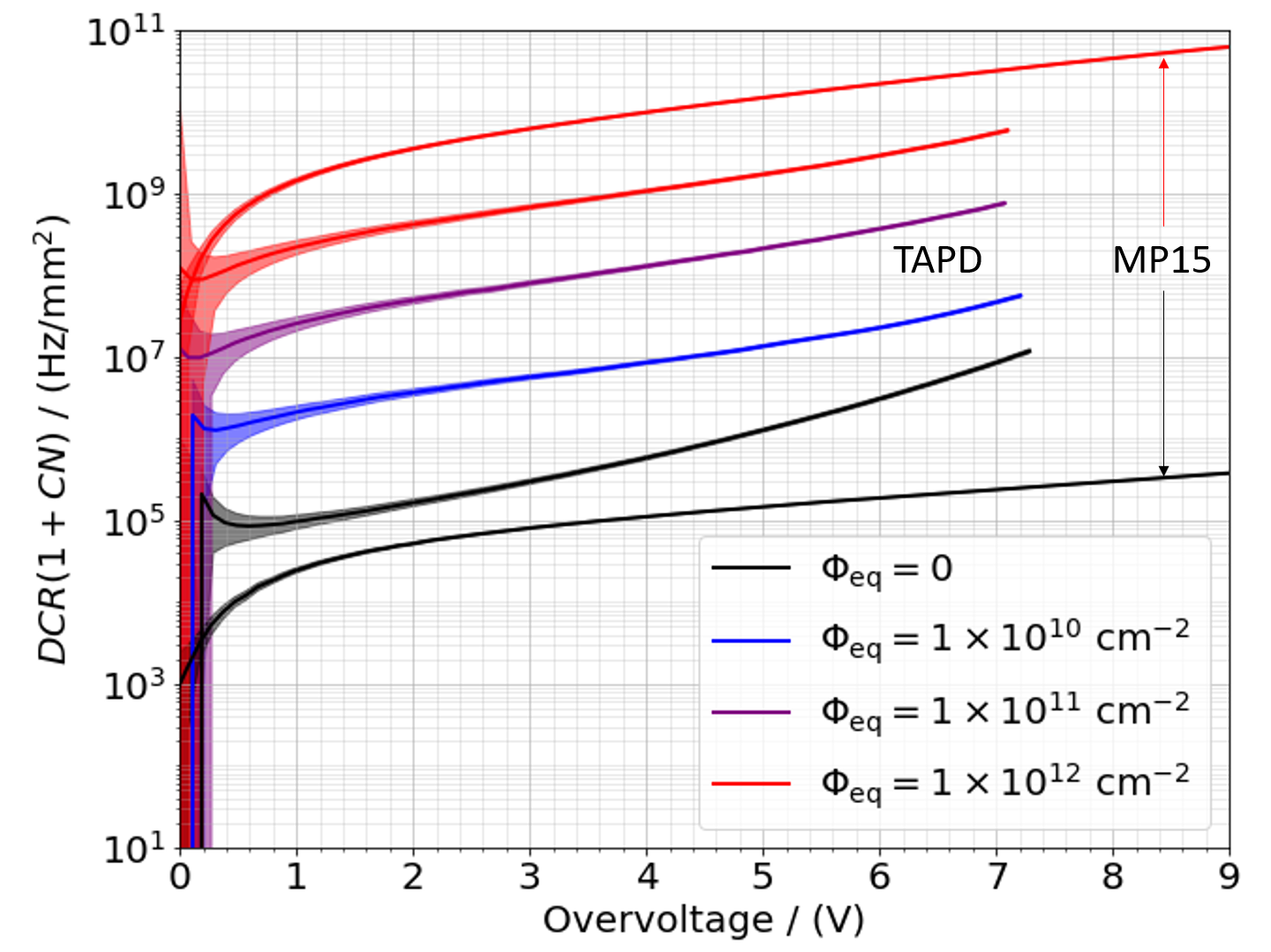}
    \caption{The dark count rate at \SI{20}{\degree C} calculated using Eq.~\ref{eq:dcr} normalized to a detector area of \SI{1}{mm^2} for the TAPD \SI{0.6}{\micro\metre} and the MP15. The overvoltage is given as $V_{\text{over}}=V_{\text{bias}}-V_{\text{BD}}$.}
    \label{fig:dcr}
        \end{minipage}
\end{figure}

The dark currents of the MP15 and the TAPD with a pillar diameter of \SI{0.6}{\micro\metre} at \SI{20}{\degree C} is shown in Fig.~\ref{fig:idark}. The increase of dark current above the breakdown is in the order of \SI{1e3}{} for the TAPD and \SI{1e5}{} for the MP15 at a fluence of \SI{1e12}{cm^{-2}}. 

The calculated dark count rates using Eq.~\ref{eq:dcr} at \SI{-30}{\degree C} and \SI{20}{\degree C} are shown in table~\ref{tab:idark} and the $DCR(V)$-behavior is displayed in Fig.~\ref{fig:dcr}. 
Comparing the DCR at an overvoltage of \SI{5}{V} at low temperatures before irradiation and after irradiation with the highest fluence yield an increase of a factor of \SI{1.3e3} for the TAPD with $d_{\text{p}}=\SI{0.6}{\micro\metre}$ and \SI{1.3e5} for the MP15. At \SI{20}{\degree C}, the increase factor is \SI{1.3e3} for the TAPD and \SI{1e5}{} for the MP15. This could be related to a smaller increase in trap-assisted band-to-band tunneling caused by radiation-introduced cluster defects in the high-field region in the TAPD compared to the MP15. 

\begin{table}[h]
    \centering
    \begin{tabular}{|c||l|l|l|l|}\hline
     & \multicolumn{4}{c|}{$\bm{DCR(1+CN)}$}\\
     & \multicolumn{2}{c|}{/ (\SI{}{MHz\per mm^2})} & \multicolumn{2}{c|}{/ (\SI{}{GHz\per mm^2})} \\ \hline
      \multirow{2}{*}{Device} &  \multicolumn{2}{c|}{$\Phi_{\text{eq}}=\SI{0}{cm^{-2}}$} &  \multicolumn{2}{c|}{$\Phi_{\text{eq}}=\SI{1e12}{cm^{-2}}$}\\\cline{2-5}
     & \multicolumn{1}{c|}{\SI{-30}{\degree C}} & \multicolumn{1}{c|}{\SI{20}{\degree C}} & \multicolumn{1}{c|}{\SI{-30}{\degree C}} & \multicolumn{1}{c|}{\SI{20}{\degree C}} \\\hline \hline
     MP15 & \SI{0.015\pm0.001}{} & \SI{0.150\pm0.003}{} & \SI{2.0\pm0.1}{} & \SI{15\pm1}{} \\ \hline
     TAPD \SI{0.6}{\micro\metre} & \SI{0.150\pm0.008}{} & \SI{1.3\pm0.1}{} & \SI{0.2\pm0.01}{} & \SI{1.7 \pm 0.1}{} \\ \hline
     TAPD \SI{0.8}{\micro\metre} & \SI{0.077\pm0.003}{} & \SI{0.73\pm0.04}{} & \SI{0.2\pm0.01}{} & \SI{2.6 \pm 0.1}{} \\ \hline
     TAPD \SI{1.0}{\micro\metre} & \SI{3.7\pm0.2}{} & \SI{92\pm10}{} & \SI{0.21\pm0.01}{} & \SI{3.4 \pm 0.1}{} \\ \hline
    \end{tabular}
    \caption{Calculated values for $DCR(1+CN)$ at an overvoltage of \SI{5}{V} using Eq.~\ref{eq:dcr} and normalized to a detector area of \SI{1}{mm^2}.}
    \label{tab:idark}
\end{table}

\section{Conclusion}
We analyzed a promising SiPM prototype before and after neutron irradiation and compared it to the MP15, a SiPM with a conventional, planar \mbox{p-n} pixel design. At a fluence of $\Phi_{\text{eq}}=\SI{1e12}{cm^{-2}}$, we found a comparable increase in quenching resistance and an increase of the breakdown voltage of the TAPD by \SI{0.16\pm0.06}{\%}. The increase in dark count rate with irradiation is in the order of \SI{1e3}{}. For comparison, the dark count rate of the MP15 increases by a factor of \SI{1e5}{} at the same temperature and overvoltage. This increased radiation hardness of the TAPD may be attributed to the geometry of the high-field region, which is roughly a factor of 3 smaller than that of the MP15.
\section*{Acknowledgements}
We acknowledge the support from BMBF via the High-D consortium. We thank Anže Jazbec  and Sebastjan Rupnik for the neutron irradiations performed at the TRIGA reactor of the Jožef Stefan Institute, Ljubljana. We thank KETEK for providing the samples and Broadcom for the useful discussion. 
This work is supported by the Deutsche Forschungsgemeinschaft (DFG, German Research Foundation) under Germany’s Excellence Strategy, EXC 2121, Quantum Universe (390833306).
 \bibliographystyle{elsarticle-num} 
\bibliography{sipm2}




\end{document}